%% file: samplepaper.tex
\documentclass[runningheads, anonymous]{llncs}
\usepackage{graphicx}
\usepackage{subcaption}
\begin{document}
\title{Your Day in Your Pocket: Complex Activity Recognition from Smartphone Accelerometers}

\author{Emma Bouton--Bessac%\orcidID{0000-1111-2222-3333} 
\and
Lakmal Meegahapola%\orcidID{1111-2222-3333-4444} 
\and
Daniel Gatica-Perez}%\orcidID{2222--3333-4444-5555}}

\authorrunning{Bouton--Bessac et al.}
\titlerunning{Complex Activity Recognition from Smartphone Accelerometers}

\institute{Idiap Reseach Institute, Switzerland \\
EPFL, Switzerland}

\maketitle              % typeset the header of the contribution
\begin{abstract}

Human Activity Recognition (HAR) enables context-aware user experiences where mobile apps can alter content and interactions depending on user activities. Hence, smartphones have become valuable for HAR as they allow large, and diversified data collection. Although previous work in HAR managed to detect simple activities (i.e., sitting, walking, running) with good accuracy using inertial sensors (i.e., accelerometer), the recognition of complex daily activities remains an open problem, specially in remote work/study settings when people are more sedentary. Moreover, understanding the everyday activities of a person can support the creation of applications that aim to support their well-being.
This paper investigates the recognition of complex activities exclusively using smartphone accelerometer data. We used a large smartphone sensing dataset collected from over 600 users in five countries during the pandemic and showed that deep learning-based, binary classification of eight complex activities (sleeping, eating, watching videos, online communication, attending a lecture, sports, shopping, studying) can be achieved with AUROC scores up to 0.76 with partially personalized models. This shows encouraging signs toward assessing complex activities only using phone accelerometer data in the post-pandemic world.

\keywords{smartphone sensing, human activity recognition, accelerometer, deep learning}
\end{abstract}
\input{1introduction}
\input{2background}
\input{3methods}

\input{4results}
\input{5discussion}
\input{6conclusion}

\bibliographystyle{splncs04}
%\newpage
\bibliography{mybibliography}
%\end{thebibliography}

\end{document}

%% file: 1introduction.tex
\section{Introduction} \label{intro}

In Human Activity Recognition (HAR), various human activities such as walking, running, sitting, [...], cooking, driving are recognized. The data can be collected from wearable sensors or accelerometer or through video frames or images \cite{HAR_a_survey}. HAR is possible thanks to sensor data from modalities such as accelerometer, gyroscope, or location \cite{DLforHARinMC} \cite{straczkiewicz2021systematic}. According to Plötz et al. \cite{DLforHARinMC}, the main challenges of HAR are the lack of data and the poor quality and labeling of the data. Recent devices like smartwatches allow for good-quality data for HAR. For example, using a smartwatch, Laput et al. \cite{sensingfinegrainedhandact} obtained high accuracies for classifying 25 complex hand activities. However, smartwatch adoption is much lower compared to smartphones, and according to Coorevits et al. \cite{riseandfallwearables}, most people tend to stop using smartwatches and wearables after six months of use.

Using smartphones for HAR seems promising given their ubiquity: more than 80\% of people own a smartphone, which could simplify data collection and increase the amount of data. Data collection can be performed on diverse populations, and continuous collection is possible. The data collected are diverse because there are numerous sensors in a smartphone, such as an accelerometer, gyroscope, light sensor, magnetic field, app usage, typing and touch events, etc. \cite{meegahapola2020smartphone}. Multiple sensing modalities also allow recognizing complex activities such as eating \cite{meegahapola2022sensing} and drinking \cite{bae2017detecting}, and even complex psychological states such as mood \cite{servia2017mobile}. Using smartphones for young adults' well-being is also increasingly popular \cite{meegahapolaWellBeing} because of the high smartphone ownership in this population. Understanding one's everyday activities can help create applications to improve mental health. Also, using data from different countries involves taking into account different cultures, people, sensor qualities, and ways to carry a smartphone (pocket, backpack, purse, etc.). Therefore, data from multiple countries should generalize better, although bringing additional challenges. Using multiple sensing modalities, while informative, could be costly in terms of battery life. Hence, there is a push towards only using low-cost inertial sensors for HAR \cite{HAR_a_review}. 

Previous work on HAR that use inertial sensors focuses on inferring relatively simple activities such as walking, sitting, climbing stairs, and sleeping \cite{betterPACusingSmartphone,HASSAN2018}. However, recognizing complex activities can be helpful in various situations, such as elderly care and patient tracking \cite{sensingfinegrainedhandact,review_HAR_apps} and for habit tracking (e.g., to help people quit smoking  \cite{sensingfinegrainedhandact}). Moreover, due to the pandemic, most people's everyday life has changed to a more sedentary lifestyle, making the HAR tasks even more challenging because the informativeness of smartphone accelerometers could be less. 

In this work, we attempt to address the research question (RQ): Can only raw accelerometer data be used to recognize complex daily activities with data collected during the pandemic (remote study setting)? In addressing this RQ, two contributions are provided:

\textbf{Contribution 1}: We examine a real-life smartphone sensing dataset that contains over 216K self-reports from 637 college students in five countries. The dataset was collected for four weeks during the pandemic. We perform a descriptive data analysis to identify the most common complex activities reported by participants.

\textbf{Contribution 2}: We define and evaluate binary inference models for eight complex daily activities: Sleeping, Eating, Studying, Attending a lecture, Online Communication and Social Media, Watching videos or TV, Sports, and Shopping, all of which represent facets of the everyday life of young adults. Using only raw accelerometer data and deep learning, we show that AUROC scores in the range of 0.51-0.62 can be achieved with population-level models, and it could be improved to AUROC scores in the range of 0.56-0.76 with hybrid models. 

To the best of our knowledge, our work contributes to understanding how the sole use of smartphone accelerometer data can be used for the inference of complex activities like the ones we study here. The pandemic context enhanced remote work and sedentary lifestyles, so it is a setting worth investigating.
The paper is organized as follows. In Section \ref{background}, the related work is presented. Then, the methods and results are explained in Section \ref{methods} and Section \ref{results} respectively. Finally, the main findings are discussed in Section \ref{discussion}, and the paper is concluded in Section \ref{conclusion}.

%% file: 2background.tex
\section{Background and related work}\label{background}

\subsection{Smartwatches and HAR}
\subsubsection{{Wearables for HAR.}}
Laput et al. \cite{sensingfinegrainedhandact} managed to capture fine-grained hand activities using smartwatches. There were 25 hand activities such as clapping, drinking, or door opening. Using Fast Fourier Transform and Convolutional Neural Networks, the method yielded 95.2\% accuracy across the 25 hand activities. This work could be used to track habits such as smoking or for eldercare monitoring systems. One disadvantage is that the user must wear the device on their active arm, whereas smartwatches are usually worn on the passive arm. HAR with smartwatches on the passive arm would be more challenging but more adapted to real life. Another challenge with smartwatches, shown by Straczkiewicz et al. \cite{placementlocation}, is that about 15.6\% people do not follow the data collection protocol regarding smartwatch placement, such as wearing the watch on the 'wrong' arm. This study shows that the data collection for HAR can be very challenging, as simply wearing a sensor on the non-ideal arm can decrease performance. These results are also valid for real-life applications: if the sensors are misplaced, the detected activity could be incorrect.

\subsubsection{{Smartphones vs. Smartwatches for HAR.}}
Raihani et al. \cite{Mohamed2018MULTILABELCF} showed that classifiers can perform as well as when the accelerometer is placed in the pocket rather than on the wrist for basic activities (sitting, walking, running). Smartphones have a practical advantage in the long run, as many users stop using their smartwatches after a few months \cite{riseandfallwearables}. Performing HAR with smartphones can be as efficient as with wearables, and more data can be used as the ownership of smartphones is higher than that of smartwatches. Furthermore, combining smartphone sensors and wrist-worn motion sensors is even more effective than only using smartwatches \cite{complexHARusingSmarphoneandWWmotionsensors}. Such work evaluated basic activities along with more complex ones like smoking, biking, or drinking coffee. The results showed that combining the sensors from the phone and the watch improves the performance by 21\%, for an overall F1 measure of 96\%. However, the work in \cite{complexHARusingSmarphoneandWWmotionsensors} was performed in a lab setting, and its application to real-life cases will probably yield lower performance.

\subsection{Smartphones and HAR}
\subsubsection{{Sensors and Features}}
Smartphone sensors can be used to infer a variety of human activities and states. For example, mood can be inferred from social interaction data \cite{moodscope}. Features like the number of SMS, emails, and apps used are fed into various machine learning models to assess user mood. The method achieved 93\% accuracy after a two-months personalized training period. Guvensan et al. \cite{NovelSegmentBasedApproach} used the smartphone's accelerometer, gyroscope, and magnetometer sensors to assess the transport mode. The method achieved 95\% accuracy with supervised learning approaches. Hassan et al. \cite{HASSAN2018} extracted features (mean, frequency skewness, average energy) from the smartphone's gyroscope and accelerometer and fed them into a Deep Belief Network. The method achieved 89.61\% accuracy on basic activities (walking, sitting, walking upstairs/downstairs) and the transitions between two activities. Wu et al. \cite{ClassificationAccuracies} used the same activities performed at different paces and collected data from the accelerometer. Their method obtained accuracies between 52-79\% for stair walking and up to 100\% for sitting; adding the gyroscope data improved the performance by 3.1-13.4\%. Della Mea et al. \cite{FeasibilityStudyonSmartphoneAccBasedRecognitionofHouseholdAct} used the smartphone's accelerometer to infer household activities such as working at the computer, ironing, or sweeping the floor. Their proposed method obtained an accuracy above 80\%, even when the phone was in the pocket. This gives initial evidence to support the hypothesis that the phone's accelerometer alone could also be used for recognizing complex activities. 

\subsubsection{{Complex Activities}}
Ranasinghe et al. \cite{review_HAR_apps} defined a complex activity as a succession of simple actions. The actions are composed of operations, which are the basic steps constituting the actions. For instance, the complex activity "Party" can be broken down into actions such as "meet with friends", "enter a bar", and "order a drink". These actions can then be broken down into operations like "push the door handle" or "grab the glass". Complex activities can include interactions with objects or individuals (such as eating, communicating online, and partying) and last longer in time. HAR can monitor the complex activities of elderly people and improve their quality of life. Healthcare monitoring applications are also an interesting field, and using only the smartphone to recognize activities is not invasive, compared to previous work that often uses body sensors \cite{wearablesensorbasedhandgesture}. Using the minimum amount of sensors allows for a spare battery and would also be more efficient memory-wise. However, it is more challenging because there will be less data, and this data can be less meaningful for some complex activities.

Our work differs from previous work regarding the inferred activities and the sensors used. We aim to infer complex activities like studying or eating, exclusively using raw accelerometer data collected in everyday life. This makes the inference challenging compared to HAR models trained with data collected in in-lab settings. Further, as mentioned in Section \ref{methods}, the dataset being collected during the COVID-19 pandemic represents a challenge because accelerometer data will likely be similar for different activities. Hence, there is a novelty in studying how complex activity recognition models perform with data collected during the pandemic. 

%% file: 3methods.tex
\section{Methods}\label{methods}

\subsection{Dataset}

The anonymized data used in this study was collected as part of a European Union Horizon 2020 Project called WeNet \cite{wenet}. The data were collected in the fall of 2020. The original study aimed to measure aspects of the diversity of university students based on social practices and related daily behaviors, combining mobile surveys and smartphone sensor data. The study was conducted at Aalborg University (Denmark), the London School of Economics (United Kingdom), the National University of Mongolia (Mongolia), Universidad Católica "Nuestra Señora de la Asunción" (Paraguay), and the University of Trento (Italy).

\begin{figure}[ht]
     \centering
     \begin{subfigure}[b]{0.45\textwidth}
         \centering
         \includegraphics[width=\textwidth]{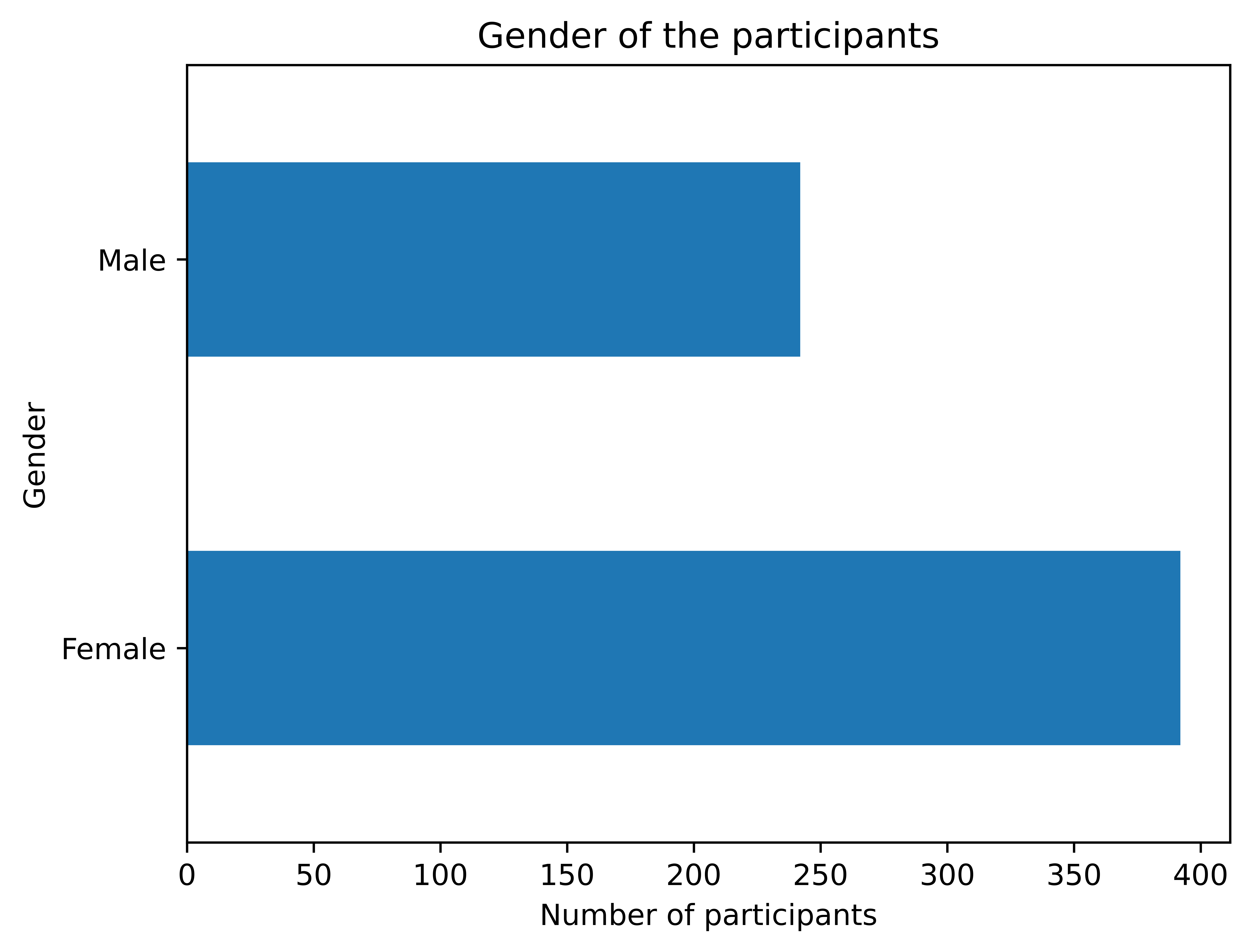}
         \caption{Gender distribution of the dataset.}
         \label{fig:gender}
     \end{subfigure}
     \hfill
     \begin{subfigure}[b]{0.45\textwidth}
         \centering
         \includegraphics[width=\textwidth]{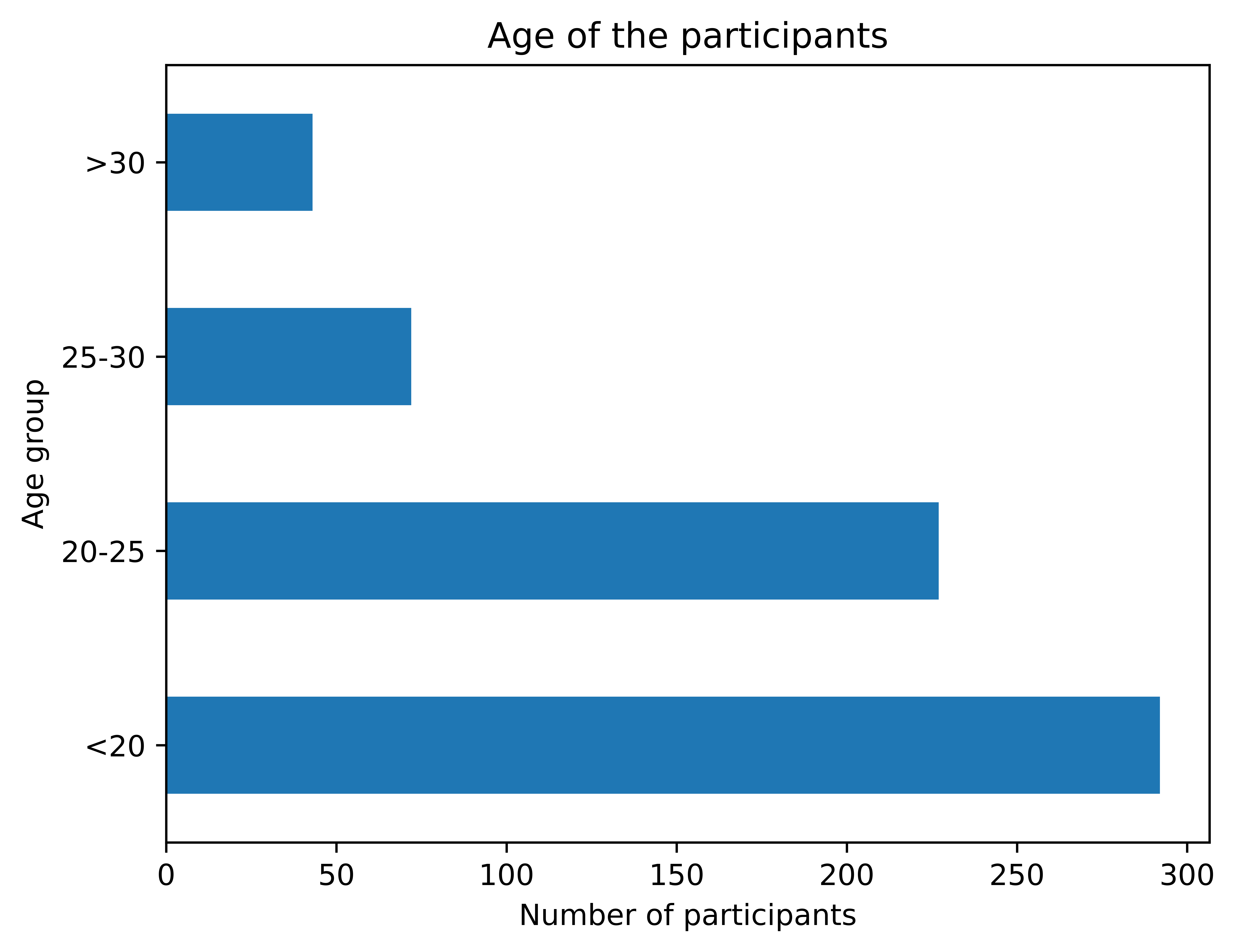}
         \caption{Age distribution of the dataset.}
         \label{fig:age}
     \end{subfigure}
        \caption{Gender and age distribution of the dataset.}
        \label{fig:genderage}
\end{figure}

A sample of volunteer students participated in a four-week data collection. The students were approached by the data collectors via an email to the entire population enrolled in the universities that took part in the survey \cite{wenet_diversity}. After having consented to the processing of their personal data, agreed to participate and have consented to be contacted along with having a smartphone version of Android 6.0 or higher, participants filled out a time diary via a mobile app  \cite{wenet_diversity}. Participants were 61\% females and average age was 22 years old (see Figure \ref{fig:genderage}). The app sent notifications every hour for the four weeks, asking the participant to complete a time diary (also referred to as self-report) to report their current activity, among other variables not used in this paper. If the participant could not answer the questionnaire, they could fill it in later (for example, when they woke up, they could indicate they have been sleeping for the past hours). The students received incentives at the end of the study. 
The activity list was defined according to previous survey work in sociology. In the meantime, the application collected data from 34 sensors, such as the accelerometer, gyroscope, battery level, app usage, etc. Here, only the raw accelerometer data will be used (other sensor data could have been used, but they were not considered as our focus here is specifically on the accelerometer data). After data pre-processing and filtering, approximately 40K self-reports were available for analysis. 

\subsection{Data preparation}

\subsubsection{Class Selection.} 
\begin{figure}[ht]
     \centering
     \begin{subfigure}[b]{0.64\textwidth}
         \centering
         \includegraphics[width=\textwidth]{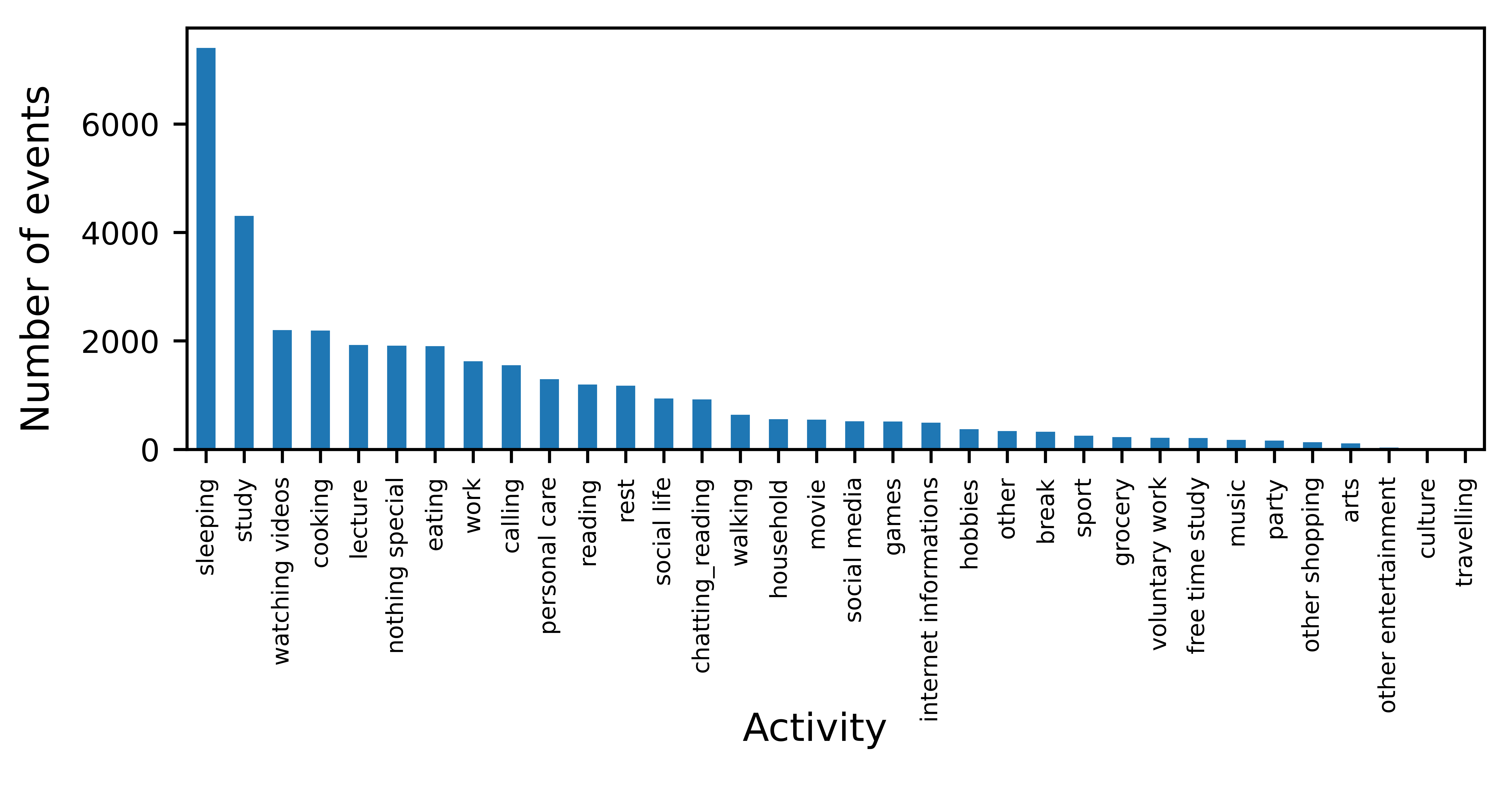}
         \caption{Number of reports per activity for the whole dataset.}
         \label{fig:reportsall}
     \end{subfigure}
     \hfill
     \begin{subfigure}[b]{0.35\textwidth}
         \centering
         \includegraphics[width=\textwidth]{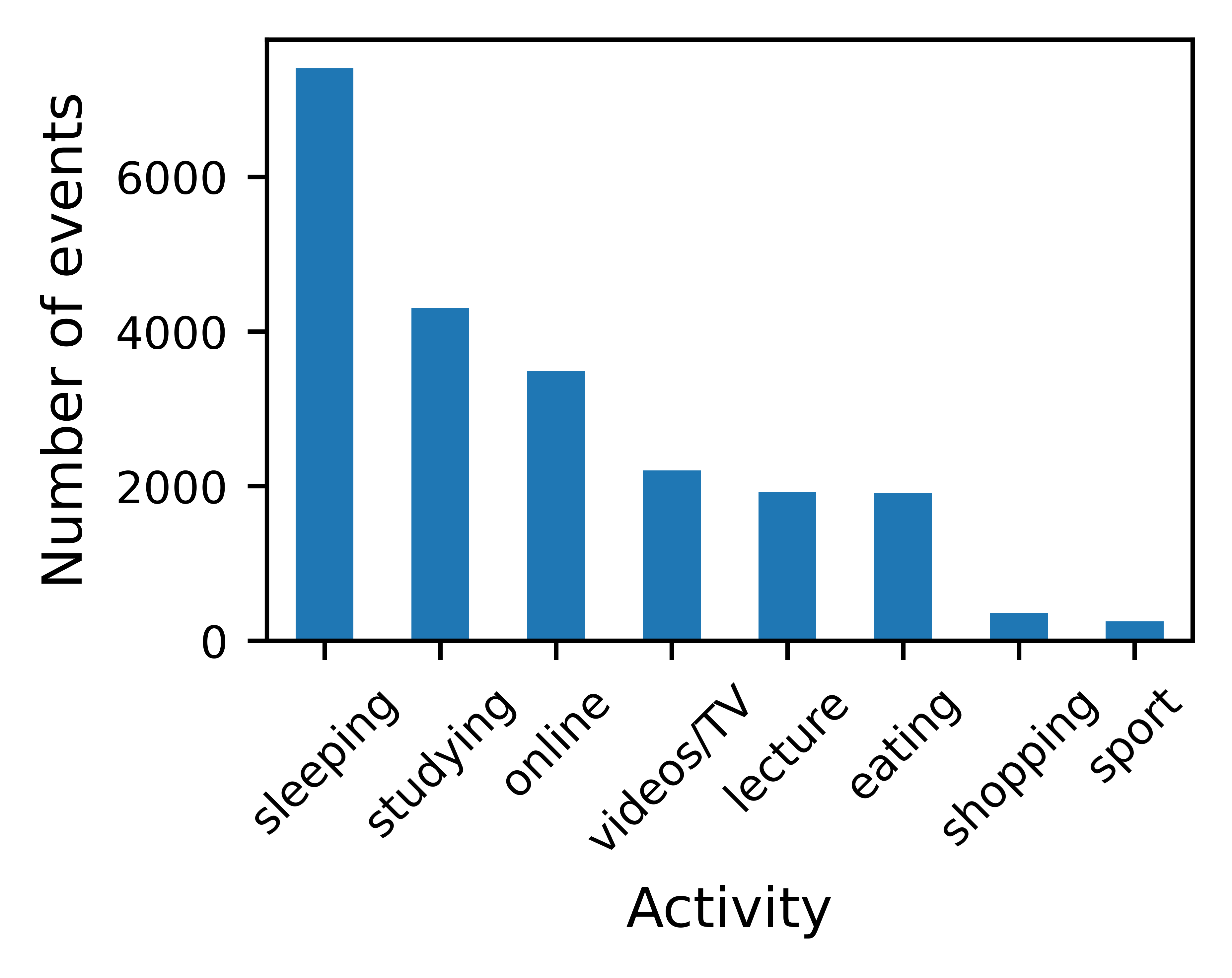}
         \caption{Number of reports per activity for the final list of activities.}
         \label{fig:reportsact}
     \end{subfigure}
        \caption{Number of reports per activity.}
        \label{fig:reportsperactivity}
\end{figure}

Figure \ref{fig:reportsall} shows the dataset's number of events per activity. The large class imbalance is not surprising, but it means that there is not enough data for all activities: {\it Travelling} counts only 19 events across all five countries, which is not enough for a model to learn. Therefore, {\it Movie, theatre, concert}, {\it Hobbies}, {\it Arts}, {\it Happy hour/drinking}, {\it Other entertainment}, {\it Entertainment Exhibit, Culture}, and {\it Travelling} have not been taken into account in this work, because of the lack of data.  Other activities were not considered because they are too broad: {\it Personal care}, {\it Games}, {\it Social life}, or {\it Voluntary work}, involve many possible complex activities. Activities like {\it Nothing special},  {\it Break}, and {\it Other} are too general and thus not interesting to infer. In addition, we decided to merge some classes: {\it Shopping} and {\it Other shopping} were merged into {\it Shopping}; while {\it Calling}, {\it Chatting/reading}, {\it Reading internet information}, and {\it Social media} were merged into {\it Online communication and social media}.

After this process, eight classes were finally kept: {\it Sleeping, Eating, Studying, Attending a lecture, Online communication and social media, Watching videos/TV, Sports, Shopping}. Their distribution is shown in Figure \ref{fig:reportsact}. These eight classes are still diverse and specific enough to allow training. In addition, they are interesting cases of the complex everyday activities of university students. Many of them directly impact health (sleeping, eating, doing sports) and some indirectly (Online communication and social media, studying), so these activities are worth inferring regarding young adults' well-being.

\subsubsection{Pre-Processing.} Previous HAR work has used time windows of 2-30 seconds to infer basic activities \cite{betterPACusingSmartphone,complexHARusingSmarphoneandWWmotionsensors}. However, the activities of interest in this work are complex, which means that they last longer \cite{review_HAR_apps}; thus, it might be harder to recognize complex activities in only a couple of seconds. Another aspect of this task is that it is unclear exactly when the participant is performing the reported activity. As they fill out the form every 30 minutes, it is unclear whether they are doing the activity during the entire 30-minute period, only before the report or only after. Therefore, one assumption was that the user performs the activity sometime during the 3 minutes before the self-report time (i.e., the timestamp when the participant completed the self-report). It was further assumed that the person did not perform the reported activity during the completion of the report itself (i.e., running while completing a report about running), so the data corresponding to this specific time has been removed. Hence, the following process was applied to the data:
\begin{enumerate}
    \item For each user's self-report, select the accelerometer data for the last 3 minutes before the report time.
    \item Remove the time during which the user fills the self-report. %(if there is no data remaining, discard the event.)
    \item Re-sample the accelerometer data to the average sampling frequency of the dataset (3.33Hz)
    \item If a report is shorter than 600 samples (3min * 60 seconds * 3.33), the data point is discarded.
\end{enumerate}

\begin{figure}[ht]
  \centering
  \begin{minipage}[b]{0.45\textwidth}
    \raisebox{.6\height}{\includegraphics[width=1.1\textwidth]{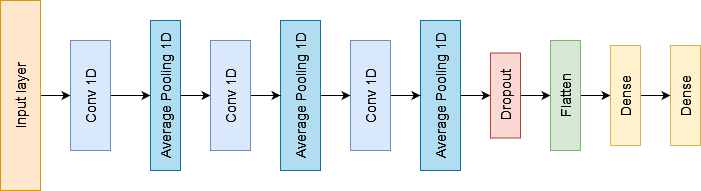}}
    \caption{Model Architecture.}
    \label{fig:model_archi}
  \end{minipage}
  \hfill
  \begin{minipage}[b]{0.45\textwidth}
    {\includegraphics[width=0.8\textwidth]{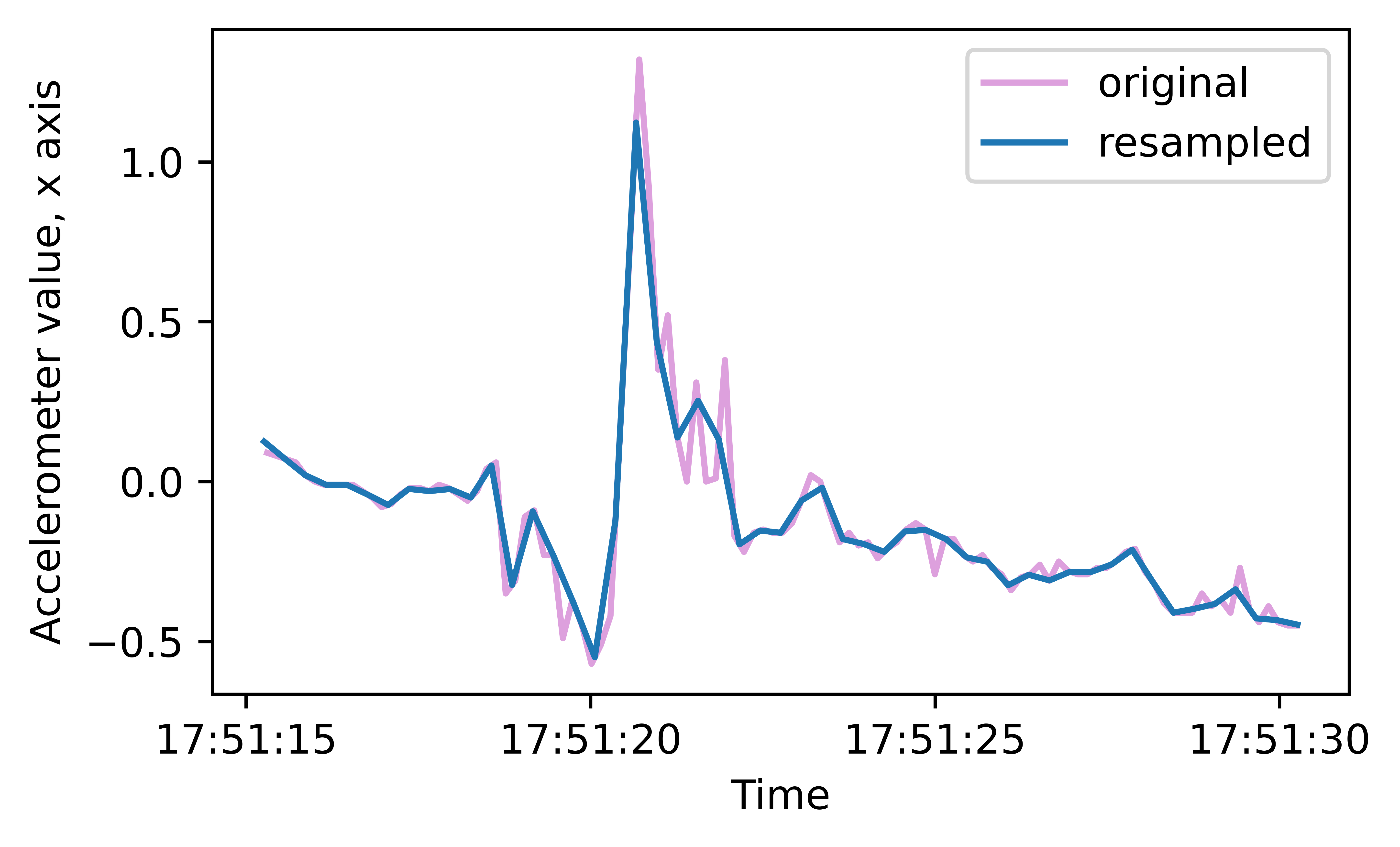}}
    \caption{"Attending lecture" accelerometer value for a Denmark participant, x axis.}
    \label{fig:resampled_denmark}
  \end{minipage}
\end{figure}

Upon inspection, we noticed significant missing data: even though the data pre-processing was the same for all five countries, many events were discarded because there needed to be accelerometer data within a 10-minute time window around the self-report time. For Mongolia and Italy, most reports were discarded because there was no data. Paraguay, Denmark, and the UK have a small number of empty reports. Mongolia and Italy are where most data was gathered, so the data loss is significant: they represent 160K reports, whereas Paraguay, Denmark, and the UK gather 31K reports. This represents a challenge because it reduces the amount of data available for deep learning. The remaining self-reports were resampled to the average sampling frequency of the dataset (see Figure \ref{fig:resampled_denmark}). The average sampling time was computed for seven users of each country and rounded to 300ms.

\subsection{Deep Learning} \label{deeplearning}

The previous section discussed the last three minutes of accelerometer data before each self-report was used. The input data has a shape of 3x600: three accelerometer axes, x, y, and z, and a data length of 600. Each report is labeled with the corresponding activity. The data is then fed to a Deep Learning model. Several architectures were implemented, using 1D Convolutional layers or LSTM layers, and the best performing model was used, as shown in Figure \ref{fig:model_archi}. 1D Convolutional layers were used because the input data is a sequence. LSTM models were explored because it is usually used to process sequences of data. It did not give better results than 1D Convolutions in this case. In addition, the Adam optimizer and binary cross-entropy were used. The performance measures used were accuracy, the area under the receiver operating characteristic curve (AUROC), and the F1 score. We reported all three metrics because accuracy makes sense in a balanced class setting. However, AUROC and F1 scores with macro averaging make sense in the imbalanced class setting because they give equal emphasis to both majority and minority classes. The evaluation was done with 10-fold cross-validation.

The binary classification was performed for each class, such as 'sleeping' and 'not sleeping'. For each training split, 60\% of samples from the positive class were randomly selected for training, 20\% for validation, and 20\% for testing. Two settings were tested, with balanced and imbalanced data. The same amount of samples were selected from the seven other classes for the balanced case. All data from the seven other classes were used for the imbalanced case. Training on an imbalanced dataset was done to determine whether niche events (events that do not occur often) can be inferred. For instance, shopping is a fraction of the week of a student, and it is interesting to see whether it could be detected.

The training was run on a population level, meaning that data is split users-wise: users can only be in one set (training users are not in validation or test). The results of this experiment will show whether complex activity recognition can perform well on new users. 
Another experiment was to split the data samples-wise: reports were randomly split into train, validation, and test.

%% file: 4results.tex
\section{Results}\label{results}

% LOUO
\begin{table*}[t]
        \small
        \centering
        \caption{Population-Level Results: AUROC score, F1-score, Accuracy and standard deviation for balanced and imbalanced datasets.}
        \resizebox{0.95\textwidth}{!}{%
        
        \begin{tabular}{l l l l l l l l l l l l l}

        & \multicolumn{6}{c}{\textbf{Balanced dataset}} & \multicolumn{6}{c}{\textbf{Imbalanced dataset}} \\

        \textbf{Activity} &
        
        \multicolumn{2}{c}{\textbf{AUROC}} &
        \multicolumn{2}{c}{\textbf{F1-score}} &
        \multicolumn{2}{c}{\textbf{Accuracy (\%)}} &
        \multicolumn{2}{c}{\textbf{AUROC}} &
        \multicolumn{2}{c}{\textbf{F1-score}} &
        \multicolumn{2}{c}{\textbf{Accuracy (\%)}} 
        \\

        & \textbf{mean} & \textbf{std} & \textbf{mean} & \textbf{std} & \textbf{mean} & \textbf{std} & \textbf{mean} & \textbf{std} & \textbf{mean} & \textbf{std} & \textbf{mean} & \textbf{std} \\
        \hline

        Sleeping & % activity name
        % ACC& % feature group
        % balanced
        
        0.62 & % AUC mean
        0.04 & % AUC std
        0.56 & % f1 mean
        0.20 &  % f1 std
        58.2 & % accuracy mean
        2.1 & % accuracy std
        % imbalanced
        
        0.62 & % AUC mean
        0.05 & % AUC std
        0.63 & % f1 mean
        0.14 & % f1 std
        92.2 & % accuracy mean
        4.2  % accuracy std
        
        \\
        Eating & % activity name
        % balanced
        
        0.51 & % AUC mean
        0.03 & % AUC std
        0.79 & % f1 mean
        0.13 & % f1 std
        50.9 & % accuracy mean
        2.1 & % accuracy std
        % imbalanced
        
        0.51 & % AUC mean
        0.03 & % AUC std
        0.01 & % f1 mean
        0.01 & % f1 std
        88.8 & % accuracy mean
        2.3 % accuracy std
        
        \\
        
        Studying & % activity name
        % balanced
        
        0.55 & % AUC mean
        0.03 & % AUC std
        0.59 & % f1 mean
        0.21 & % f1 std
        53.0 & % accuracy mean
        2.2 & % accuracy std
        %imbalanced
        
        0.57 & % AUC mean
        0.04 & % AUC std
        0.11 & % f1 mean
        0.09 & % f1 std
        74.3& % accuracy mean
        4.5 % accuracy std
        
        \\
        
        Attending lecture & % activity name
        % ACC& % feature group
        
        0.52 & % AUC mean
        0.03 & % AUC std
        0.68 & % f1 mean
        0.09 & % f1 std 
        51.6 & % accuracy mean
        2.3 & % accuracy std
        %imbalanced
        
        0.51 & % AUC mean
        0.00 & % AUC std
        0.01 & % f1 mean
        0.01 & % f1 std
        89.4 & % accuracy mean
        1.9  % accuracy std
        \\
        
        Online communication & % activity name
        % ACC& % feature group
        
        0.56 & % AUC mean
        0.03 & % AUC std
        0.64 & % f1 mean
        0.21 & % f1 std 
        55.1 & % accuracy mean
        2.5 & % accuracy std
        %imbalanced
        
        0.57 & % AUC mean
        0.02 & % AUC std
        0.03 & % f1 mean
        0.02 & % f1 std
        84.7 & % accuracy mean
        1.8 % accuracy std
        \\
        
        Watching videos/TV & % activity name
        % ACC& % feature group
        
        0.54 & % AUC mean
        0.04 & % AUC std
        0.51 & % f1 mean
        0.19 & % f1 std 
        53.0 & % accuracy mean
        2.5 & % accuracy std
        %imbalanced
        
        0.56 & % AUC mean
        0.03 & % AUC std
        0.04 &% f1 mean
        0.03 & % f1 std
        85.3 & % accuracy mean
        2.2 % accuracy std
        \\
        
        Sport & % activity name
        % ACC& % feature group
        
        0.52 & % AUC mean
        0.07 & % AUC std
        0.59 & % f1 mean
        0.23 & % f1 std 
        50.9 & % accuracy mean
        7.4 & % accuracy std
        %imbalanced
        
        0.52 & % AUC mean
        0.06 & % AUC std
        0.00 & % f1 mean
        0.04 & % f1 std
        97.7 & % accuracy mean
        1.0 % accuracy std
        \\
        
        Shopping & % activity name
        % ACC& % feature group
        
        0.57 & % AUC mean
        0.06 & % AUC std
        0.58 & % f1 mean
        0.23 & % f1 std
        55.3 & % accuracy mean
        5.2 & % accuracy std
        %imbalanced
        
        0.48 & % AUC mean
        0.05 & % AUC std
        0.00 & % f1 mean
        0.00 & % f1 std
        97.3 & % accuracy mean
        0.5 % accuracy std
        \\

        \hline 
        
        \end{tabular}}
        \label{tab:LOUO}

\end{table*}
% LORO
\begin{table*}[t]
        \small
        \centering
        \caption{Hybrid Results: AUROC score, F1-score, Accuracy and standard deviation for balanced and imbalanced datasets.}
        \resizebox{.95\textwidth}{!}{%
        
        \begin{tabular}{l l l l l l l l l l l l l}

        & \multicolumn{6}{c}{\textbf{Balanced dataset}} & \multicolumn{6}{c}{\textbf{Imbalanced dataset}} \\

        \textbf{Activity} &
        
        \multicolumn{2}{c}{\textbf{AUROC}} &
        \multicolumn{2}{c}{\textbf{F1-score}} &
        \multicolumn{2}{c}{\textbf{Accuracy (\%)}} &
        \multicolumn{2}{c}{\textbf{AUROC}} &
        \multicolumn{2}{c}{\textbf{F1-score}} &
        \multicolumn{2}{c}{\textbf{Accuracy (\%)}} 
        \\

        & \textbf{mean} & \textbf{std} & \textbf{mean} & \textbf{std} & \textbf{mean} & \textbf{std} & \textbf{mean} & \textbf{std} & \textbf{mean} & \textbf{std} & \textbf{mean} & \textbf{std} \\
        \hline

        Sleeping & % activity name
        % ACC& % feature group
        % balanced
        
        0.76 & % AUC mean
        0.01 & % AUC std
        0.57 & % f1 mean
        0.06 &  % f1 std
        69.6 & % accuracy mean
        0.9 & % accuracy std
        % imbalanced
        0.72 & % AUC mean
        0.03 & % AUC std
        0.61 & % f1 mean
        0.06 &% f1 std
        66.4 & % accuracy mean
        2.8 % accuracy std
        
        \\
        Eating & % activity name
        % balanced
        0.57 & % AUC mean
        0.04 & % AUC std
        0.59 & % f1 mean
        0.30 & % f1 std
        55.3 & % accuracy mean
        2.6 & % accuracy std
        % imbalanced
        0.56 & % AUC mean
        0.04 & % AUC std
        0.71 & % f1 mean
        0.26 & % f1 std
        54.3 & % accuracy mean
        3.9 % accuracy std
        
        \\
        
        Studying & % activity name
        % balanced
        0.66 & % AUC mean
        0.013 & % AUC std
        0.68 & % f1 mean
        0.05 & % f1 std
        61.5 & % accuracy mean
        0.9 & % accuracy std
        %imbalanced
        0.67 & % AUC mean
        0.00 & % AUC std
        0.68 & % f1 mean
        0.06 & % f1 std
        61.5 & % accuracy mean
        0.9 % accuracy std
        
        \\
        
        Attending lecture & % activity name
        % ACC& % feature group
        0.62 & % AUC mean
        0.04 & % AUC std
        0.69 & % f1 mean
        0.22 & % f1 std 
        58.5 & % accuracy mean
        3.9 & % accuracy std
        %imbalanced
        0.60 & % AUC mean
        0.04 & % AUC std
        0.70 & % f1 mean
        0.22 & % f1 std
        57.8 & % accuracy mean
        3.1  % accuracy std
        \\
        
        Online communication & % activity name
        % ACC& % feature group
        0.60 & % AUC mean
        0.04 & % AUC std
        0.73 & % f1 mean
        0.11 & % f1 std 
        57.4 & % accuracy mean
        2.7 & % accuracy std
        %imbalanced
        0.61 & % AUC mean
        0.01 & % AUC std
        0.71 & % f1 mean
        0.03 & % f1 std
        58.1 & % accuracy mean
        1.3  % accuracy std
        \\
        
        Watching videos/TV & % activity name
        % ACC& % feature group
        0.56 & % AUC mean
        0.05 & % AUC std
        0.72 & % f1 mean
        0.22 & % f1 std 
        53.2 & % accuracy mean
        4.4 & % accuracy std
        %imbalanced
        0.55 & % AUC mean
        0.05 & % AUC std
        0.58 & % f1 mean
        0.07 & % f1 std
        54.2 & % accuracy mean
        4.4 % accuracy std
        \\
        
        Sport & % activity name
        % ACC& % feature group
        0.58 & % AUC mean
        0.04 & % AUC std
        0.55 & % f1 mean
        0.23 & % f1 std 
        55.6 & % accuracy mean
        3.5 & % accuracy std
        %imbalanced
        0.59 & % AUC mean
        0.03 & % AUC std
        0.72 & % f1 mean
        0.01 & % f1 std
        56.5 & % accuracy mean
        3.6 % accuracy std
        \\
        
        Shopping & % activity name
        % ACC& % feature group
        0.61 & % AUC mean
        0.06 & % AUC std
        0.54 & % f1 mean
        0.15 & % f1 std
        56.8 & % accuracy mean
        4.3 & % accuracy std
        %imbalanced
        0.63 & % AUC mean
        0.04 & % AUC std
        0.57 & % f1 mean
        0.07 & % f1 std
        59.4 & % accuracy mean
        0.1  % accuracy std
        \\

        \hline 
        
        \end{tabular}}
        \label{tab:LORO}

\end{table*}

\subsection{Population Level Model.}
The results for the population-level model can be seen on Table \ref{tab:LOUO}.

\subsubsection{Balanced Dataset.}
This could be considered as the base-case accuracy without any personalization. Hence, results indicate that it is likely every user's smartphone usage is very different, and the model does not perform well on a new user and needs personalizing. Sleeping has the highest AUROC score of 0.62. The reason could be that it is the most significant class, so there is more data to train the model. Sport has lower performance than expected, possibly because different users have different accelerometer data when engaging in sports. It could also be due to home training (in the COVID-19 context) or the users not keeping their phones in their pockets while training. Therefore, the high activity levels would not be collected. The metrics could be higher after a personalized training period. Online communication has surprisingly high metrics since it regroups four different activities. One would have expected it to yield lower metrics, but as all activities are phone-related, it makes sense.

\subsubsection{\textbf{Imbalanced Dataset}}
Here, the accuracy is not representative of the model's performance. The F1 score is very low for all activities except sleeping. The explanation is that given the class imbalance, the model always predicts the negative class and leads to a high accuracy and a low F1 score. Shopping and sport represent no more than 5\% of the dataset each, so if the model only predicts the negative class, it will lead to an accuracy of more than 95\% each, which is the case and explains the low F1 score. Only sleeping yields high metrics because it is the biggest class, so the model had more data to train. Niche events are not well recognized using the population-level approach. For shopping, the AUROC score is lower than 0.5, meaning that the model is inverting the classes in some cases.

\subsection{Hybrid model}
The metrics for the hybrid model can be seen on Table \ref{tab:LORO}.

\subsubsection{Balanced dataset.}
'Sleeping' has the best results, with an AUROC score of 0.76 for the reasons mentioned above. Also, the smartphone's activity when a person is sleeping is easy to recognize: the smartphone is probably placed on the bedside table for the night. Watching videos/TV and Online communication obtain a high F1 score. "Eating" 's metrics are low, which can be explained by the fact that people eating with their smartphones can behave differently (e.g., putting it away when eating with people, watching something, chatting, etc.). One would expect good results for "sport", but the AUROC score is only 0.58. The low metrics for Watching videos/TV can be explained by the difference between watching videos and TV. One can switch on the TV in the background and do something else on their phone (resulting in a different accelerometer activity), whereas watching a video on their phone means the attention is more focused on the phone, and the resulting accelerometer data can be the one of a phone standing still.

\subsubsection{Imbalanced Dataset.}
As mentioned in Section \ref{deeplearning}, the training was done on an imbalanced dataset to see if niche events could be recognized. 
The F1 score is generally higher for this case than for the balanced dataset. However, the AUROC score is similar on the balanced and imbalanced data.

%% file: 5discussion.tex
\section{Discussion} \label{discussion}

In the original dataset, there were 34 different activities. However, most of them were dropped because of a lack of data from the original dataset while keeping eight informative and representative activities of the life of a student. These activities also have an impact on health and can help understand the life of a student and ultimately support their well-being via applications. Using only the accelerometer data for human activity recognition is challenging because of the complexity of the activities and the COVID situation at the data collection time: the activities are more likely to result in similar accelerometer data. The resulting AUROC and F1 scores are reasonable, given the challenge. 
It was noticed that a hybrid model performs better than a population-level one and that niche events are poorly recognized (imbalanced dataset case). While the results are relatively low for binary classification, the settings must be kept in mind: the data was collected in five countries, which induces a mix of people, sensor quality, and ways of using a phone. This multi-country approach should generalize well and calls for additional studies for multi-country data. Further, studies could evaluate the data quality per country and sensor to obtain more details. The original dataset was collected in real-life conditions, which also impacts the quality of data and the performance of models, but represents real conditions in which the model would be used. As mentioned, every smartphone has different components (especially in different countries/continents), and using only the accelerometer is relevant because this sensor is cheap, and most smartphones contain built-in accelerometers. Using multiple sensors was not the focus of this paper as the goal was to use the accelerometer data only. Moreover, there is not enough gyroscope or magnetometer data in the original dataset to train a model. Also, using only one sensor would spare battery life and would have a minimal impact on a smartphone's performance in the case of a health monitoring application.

%% file: 6conclusion.tex
\section{Conclusion} \label{conclusion}

In this work, raw accelerometer data from smartphones were fed into Deep Learning models to infer complex daily activities. Binary classification led to reasonable AUROC scores in the range of 0.51-0.62 with population-level models (non-personalized) and 0.56-0.76 with hybrid models (partially personalized) for eight complex activities. This work shows that it is possible to infer complex activities using only the smartphone's accelerometer and can be a baseline for a multi-country approach of Human Activity Recognition for the well-being of young adults.

\section*{Acknowledgement}
This work was funded by the European Union’s Horizon 2020
WeNet project, under grant agreement 823783. We also thank Darshana Rathnayake (Singapore Management University, Singapore) for discussions.

%% file: samplepaper.bbl
\begin{thebibliography}{10}
\providecommand{\url}[1]{\texttt{#1}}
\providecommand{\urlprefix}{URL }
\providecommand{\doi}[1]{https://doi.org/#1}

\bibitem{wenet_diversity}
Final model of diversity,
  \url{https://www.internetofus.eu/wp-content/uploads/sites/38/2021/03/D1.3-Final-model-of-diversity.pdf}

\bibitem{HAR_a_review}
Ann, O.C., Theng, L.B.: Human activity recognition: A review. In: 2014 IEEE
  International Conference on Control System, Computing and Engineering (ICCSCE
  2014). pp. 389--393 (2014). \doi{10.1109/ICCSCE.2014.7072750}

\bibitem{betterPACusingSmartphone}
Arif, M., Bilal, M., Kattan, A., Ahamed, S.I.: Better physical activity
  classification using smartphone acceleration sensor. Journal of Medical
  Systems  (2014), \url{https://doi.org/10.1007/s10916-014-0095-0}

\bibitem{bae2017detecting}
Bae, S., Ferreira, D., Suffoletto, B., Puyana, J.C., Kurtz, R., Chung, T., Dey,
  A.K.: Detecting drinking episodes in young adults using smartphone-based
  sensors. Proceedings of the ACM on interactive, mobile, wearable and
  ubiquitous technologies  \textbf{1}(2),  1--36 (2017)

\bibitem{riseandfallwearables}
Coorevits, L., Coenen, T.: The rise and fall of wearable fitness trackers (08
  2016)

\bibitem{wenet}
Giunchiglia, F., Bison, I., Busso, M., Chenu-Abente, R., Rodas, M., Zeni, M.,
  Gunel, C., Veltri, G., de~Götzen, A., Kun, P., Ganbold, A., Chagnaa, A.,
  Gaskell, G., Stares, S., Bidoglia, M., Cernuzzi, L., Hume, A., Zarza, J.L.,
  Meegahapola, L., Gatica-Perez, D.: A worldwide diversity pilot on daily
  routines and social practices (2020). University of Trento Technical Report.
  No. \#DISI-2001-DS-0 (23),  36--44 (Apr 2021)

\bibitem{NovelSegmentBasedApproach}
Guvensan, M.A., Dusun, B., Can, B., Turkmen, H.I.: A novel segment-based
  approach for improving classification performance of transport mode
  detection. Sensors  \textbf{18}(1) (2018). \doi{10.3390/s18010087},
  \url{https://www.mdpi.com/1424-8220/18/1/87}

\bibitem{HASSAN2018}
Hassan, M.M., Uddin, M.Z., Mohamed, A., Almogren, A.: A robust human activity
  recognition system using smartphone sensors and deep learning. Future
  Generation Computer Systems  \textbf{81},  307--313 (2018).
  \doi{https://doi.org/10.1016/j.future.2017.11.029},
  \url{https://www.sciencedirect.com/science/article/pii/S0167739X17317351}

\bibitem{HAR_a_survey}
Jobanputra, C., Bavishi, J., Doshi, N.: Human activity recognition: A survey.
  Procedia Computer Science  \textbf{155},  698--703 (2019).
  \doi{https://doi.org/10.1016/j.procs.2019.08.100},
  \url{https://www.sciencedirect.com/science/article/pii/S1877050919310166},
  the 16th International Conference on Mobile Systems and Pervasive Computing
  (MobiSPC 2019),The 14th International Conference on Future Networks and
  Communications (FNC-2019),The 9th International Conference on Sustainable
  Energy Information Technology

\bibitem{sensingfinegrainedhandact}
Laput, G., Harrison, C.: Sensing Fine-Grained Hand Activity with Smartwatches,
  p. 1–13. Association for Computing Machinery, New York, NY, USA (2019),
  \url{https://doi.org/10.1145/3290605.3300568}

\bibitem{moodscope}
Likamwa, R., Liu, Y., Lane, N., Zhong, L.: Moodscope: Building a mood sensor
  from smartphone usage patterns (06 2013). \doi{10.1145/2462456.2464449}

\bibitem{FeasibilityStudyonSmartphoneAccBasedRecognitionofHouseholdAct}
Mea, V.D., Quattrin, O., Parpinel, M.: A feasibility study on smartphone
  accelerometer-based recognition of household activities and influence of
  smartphone position. Informatics for Health and Social Care  \textbf{42}(4),
  321--334 (2017). \doi{10.1080/17538157.2016.1255214},
  \url{https://doi.org/10.1080/17538157.2016.1255214}, pMID: 28005434

\bibitem{meegahapola2022sensing}
Meegahapola, L., Bangamuarachchi, W., Chamantha, A., Ruiz-Correa, S., Perera,
  I., Gatica-Perez, D.: Sensing eating events in context: A smartphone-only
  approach. IEEE Access  \textbf{10}(ARTICLE) (2022)

\bibitem{meegahapola2020smartphone}
Meegahapola, L., Gatica-Perez, D.: Smartphone sensing for the well-being of
  young adults: A review. IEEE Access  \textbf{9},  3374--3399 (2020)

\bibitem{meegahapolaWellBeing}
Meegahapola, L., Gatica-Perez, D.: Smartphone sensing for the well-being of
  young adults: A review. IEEE Access  \textbf{9},  3374--3399 (2021).
  \doi{10.1109/ACCESS.2020.3045935}

\bibitem{Mohamed2018MULTILABELCF}
Mohamed, R., Zainudin, M.N.S., Perumal, T., Mustapha, N.: Multi-label
  classification for physical activity recognition from various accelerometer
  sensor positions (2018)

\bibitem{DLforHARinMC}
Plötz, T., Guan, Y.: Deep learning for human activity recognition in mobile
  computing. Computer  \textbf{51}(5),  50--59 (2018).
  \doi{10.1109/MC.2018.2381112}

\bibitem{review_HAR_apps}
Ranasinghe, S., Machot, F.A., Mayr, H.C.: A review on applications of activity
  recognition systems with regard to performance and evaluation. International
  Journal of Distributed Sensor Networks  \textbf{12}(8),  1550147716665520
  (2016). \doi{10.1177/1550147716665520},
  \url{https://doi.org/10.1177/1550147716665520}

\bibitem{servia2017mobile}
Servia-Rodr{\'\i}guez, S., Rachuri, K.K., Mascolo, C., Rentfrow, P.J., Lathia,
  N., Sandstrom, G.M.: Mobile sensing at the service of mental well-being: a
  large-scale longitudinal study. In: Proceedings of the 26th international
  conference on world wide web. pp. 103--112 (2017)

\bibitem{complexHARusingSmarphoneandWWmotionsensors}
Shoaib, M., Bosch, S., Incel, O.D., Scholten, H., Havinga, P.J.M.: Complex
  human activity recognition using smartphone and wrist-worn motion sensors.
  Sensors  \textbf{16}(4) (2016). \doi{10.3390/s16040426},
  \url{https://www.mdpi.com/1424-8220/16/4/426}

\bibitem{straczkiewicz2021systematic}
Straczkiewicz, M., James, P., Onnela, J.P.: A systematic review of
  smartphone-based human activity recognition for health research (2021)

\bibitem{placementlocation}
Straczkiewicz, M., Glynn, N., Harezlak, J.: On placement, location and
  orientation of wrist-worn tri-axial accelerometers during free-living
  measurements (05 2019). \doi{10.3390/s19092095}

\bibitem{ClassificationAccuracies}
Wu, W., Dasgupta, S., Ramirez, E.E., Peterson, C., Norman, G.J.: Classification
  accuracies of physical activities using smartphone motion sensors. J Med
  Internet Res  \textbf{14}(5), ~e130 (Oct 2012). \doi{10.2196/jmir.2208},
  \url{http://www.jmir.org/2012/5/e130/}

\bibitem{wearablesensorbasedhandgesture}
Zhu, C., Sheng, W.: Wearable sensor-based hand gesture and daily activity
  recognition for robot-assisted living. IEEE Transactions on Systems, Man, and
  Cybernetics - Part A: Systems and Humans  \textbf{41}(3),  569--573 (2011).
  \doi{10.1109/TSMCA.2010.2093883}

\end{thebibliography}
